\documentclass[%
 reprint,
 amsmath,amssymb,
 aps,
]{revtex4-2}

\usepackage{graphicx}
\usepackage{dcolumn}
\usepackage{bm}

\usepackage{hyperref}

\newcommand{\six}[6]{\left(\begin{array}{ccc}
									{#1}& {#2}& {#3}\\
									{#4}& {#5}& {#6} \\
\end{array}\right)}

\newcommand{\ath}{a_{\ell_1 m_1} a_{\ell_2 m_2} a_{\ell_3 m_3}}

\newcommand{\athrprime}{a_{\ell_1' m_1'}(r_1') a_{\ell_2' m_2'}(r_2') a_{\ell_3' m_3'}(r_3')}

\newcommand{\afrprime}{a_{\ell_1' m_1'}(r_1') a_{\ell_2' m_2'}(r_2') a_{\ell_3' m_3'}(r_3')  a_{\ell_4' m_4'}(r_4')}

\begin{document}

\preprint{APS/123-QED}

\title{Algorithm to Produce a Density Field with Given Two, Three, and Four-Point Correlation Functions}
\thanks{zslepian@ufl.edu}%

\author{Zachary Slepian}
 \affiliation{Department of Astronomy, University of Florida, Gainesville, FL 32601}

\date{\today}

\begin{abstract}
Here we show how to produce a 3D density field with a given set of higher-order correlation functions. Our algorithm enables producing any desired two-point, three-point, and four-point functions, including odd-parity for the latter. We note that this algorithm produces the desired correlations about a set of ``primary'' points, matched to how the spherical-harmonic-based algorithms ENCORE and CADENZA measure them. These ``primary points'' must be used as those around which the correlation functions are measured. We also generalize the algorithm to i) $N$-point correlations with $N>4$, ii) dimensions other than 3, and iii) beyond scalar quantities. This algorithm should find use in verifying analysis pipelines for higher-order statistics in upcoming galaxy redshift surveys such as DESI, Euclid, Roman, and Spherex, as well as intensity mapping. In particular it may be helpful in searches for parity violation in the 4PCF of these samples, for which producing initial conditions for N-body simulations is both costly and highly model-dependent at present, and so alternative methods such as that developed here are desirable.
\end{abstract}

\maketitle


\section{\label{sec:intro} Introduction}
For a long time, it has been known how to realize a density field with a given pair correlation, either power spectrum or 2PCF \cite{hr}. This has been useful in 3D large-scale structure (LSS) analyses as well as in CMB analyses. In the CMB domain, it has also been known how to realize a given bispectrum (triple correlator of CMB temperature anisotropies) \cite{sz}. Furthermore, for LSS, it has been known how to realize certain specific bispectra, such as those produced by forms of primordial non-Gaussianity during inflation. 

Yet, a general solution to the problem of realizing arbitrary NPCFs for a 3D density field has yet to be presented. This, if done, would be quite useful in testing higher-order correlation analysis pipelines, as is indeed desirable given the wealth of upcoming 3D LSS data.

It is not quite so problematic to realize a density with a given 2PCF and 3PCF, as this can simply be done using N-body simulations. However, even here, the simulations are expensive, and each gives only a realization of the true desired NPCF. Furthermore, non-standard physics, such as parity violation in the 4PCF, is extremely expensive to simulate, and since it is very model-dependent, one must simulate many models, each at high expense.

On the observational side, surveys such as DESI have a density-dependent fiber-assignment effect, \textit{e.g.} \cite{pinol, burden, bianchi}. This could adversely effect studies of higher-order correlations. While some of this can be solved by running fiber assignment on N-body simulations, the currently-standard scheme for fixing fiber assignment errors (probability inverse pair weighting) may require a prohibitively large number of simulations if it is to be extended to the 3PCF and beyond. Thus, producing simulations with a given 3PCF, \textit{etc.} without the high expense of performing N-body runs would perhaps bring within reach the number of realizations needed for this. 

In this work, we present a method to realize any desired set of NPCFs about a given primary galaxy at $\vec{x}$. This may be done about many primaries so that the spheres of galaxies around each out to the maximal radius to which one wishes to measure the NPCF do not overlap. One could then simply run an NPCF algorithm such as \cite{encore} by flagging only the primaries desired as those to be used. This would be a suitable setup for testing the effects outlined above. The problem of optimal sphere-packing in 3D has also been solved, so one could optimally pack as many primaries as possible into a given volume.

This work is structured as follows. In Sec. \ref{sec:cmb} we review the CMB bispectrum approach; in Sec. \ref{sec:4pcf} we extend it to the 3D LSS 4PCF, and in Sec. \ref{sec:3pcf} to the 3PCF. In Sec. \ref{sec:factor} we show how the fact that many of the desired NPCF models are factorizable can be exploited to enable a fast realization scheme. We show that even seemingly non-factorizable models can be approximated by integrals with factorizable integrands, which are then approximated as Riemann sums (hence, a sum of factorized terms) to enable the same acceleration. In Sec. \ref{sec:2pcf} we show how to include the 2PCF in our set of correlations, and that one can realize any desired 2PCF, 3PCF, and 4PCF simultaneously. We discuss generalization to higher-order correlations in Sec. \ref{sec:npcf}, and to higher dimensions of space and arbitrary spatial geometry in Sec. \ref{sec:dpcf}. In Sec. \ref{sec:sphere}  we show how sphere-packing can be used to maximize the number of ``primary'' points in a given volume around which we realize the desired correlations. We conclude in Sec. \ref{sec:concs}.

\section{\label{sec:cmb} CMB Maps with a Given Power Spectrum and Bispectrum}
\subsection{CMB correlators}
\label{subsec:cmb_corr}
In the CMB context, one decomposes the temperature fluctuations around the mean as 
\begin{align}
    \frac{\Delta T (\hat{n})}{\bar{T}} = \sum_{\ell m} a_{\ell m} Y_{\ell m}(\hat{n}),
    \label{eq:cmb_decomp}
\end{align}
with $\Delta T  (\hat{n}) \equiv T(\hat{n}) - \bar{T}$, $\hat{n}$ the line of sight from the observer to the CMB, and $\bar{T}$ the mean CMB temperature. One may then compute correlators of the $a_{\ell m}$. The CMB fluctuations have statistically isotropic correlations; that is, the pair correlation, triple correlations, \textit{etc.}, are the same, on average, in any direction one looks, depending only on the relative direction vectors between the pair or triple, \textit{etc.} of CMB points. This statistical isotropy means that the correlators must be invariant under rotations about the observer.

We have the CMB power spectrum
\begin{align}
    \left<a_{\ell m} a_{\ell' m'}^* \right> = C_{\ell} \delta^{\rm K}_{\ell \ell'} \delta^{\rm K}_{m m'}.
\end{align}
The $\delta^{\rm K}_{ij}$ is the Kronecker delta, unity when its subscripts are equal and zero otherwise. There are only correlations at equal $\ell$ because of rotation invariance about the observer; unequal $\ell$ correlators vanish under rotation averaging due to orthogonality of the spherical harmonics. Similarly, the correlation also requires equal $m$, due to azimuthal symmetry under rotations around the line of sight. Furthermore, the power spectrum must be $m$-independent since $m$ is the $z$-component of $\ell$, and hence dependent on choice of $z$ axis, which would not be rotation invariant about the observer.

The CMB bispectrum $B_{\ell_1 \ell_2 \ell_3}$ \cite{spergel-bi-1, spergel-bi-2, bartolo} is defined by
\begin{align}
    \left<a_{\ell_1 m_1} a_{\ell_2 m_2} a_{\ell_3 m_3} \right> = B_{\ell_1 \ell_2 \ell_3} \six{\ell_1}{\ell_2}{\ell_3}{m_1}{m_2}{m_3},
    \label{eq:B}
\end{align}
where the matrix is a Wigner 3-$j$ symbol \cite{nist}. Given rotation-invariance about the observer, the Wigner-Eckart theorem \cite{rose} implies that the CMB bispectrum must be proportional to a 3-$j$ symbol \footnote{3-$j$ symbols are related to Clebsch-Gordan coefficients by a scaling and a phase, \textit{e.g.} \url{https://mathworld.wolfram.com/Clebsch-GordanCoefficient.html} Eq. 7.}. 

One conventionally (\textit{e.g.} \cite{bartolo}, discussion between their Eqs. 458 and 459) defines the reduced CMB bispectrum $b_{\ell_1 \ell_2 \ell_3}$ via 
\begin{align}
     B_{\ell_1 \ell_2 \ell_3} = \mathcal{C}_{\ell_1 \ell_2 \ell_3} \six{\ell_1}{\ell_2}{\ell_3}{0}{0}{0} b_{\ell_1 \ell_2 \ell_3},
     \label{eq:reduced_b}
\end{align}
with 
\begin{align}
    \mathcal{C}_{\ell_1 \ell_2 \ell_3} \equiv \sqrt{\frac{(2\ell_1 + 1) (2\ell_2 + 1) (2\ell_3 + 1)}{4\pi}}.
    \label{eq:calC}
\end{align}
Combining Eqs. \ref{eq:B} and \ref{eq:reduced_b} shows that 
\begin{align}
    \left<a_{\ell_1 m_1} a_{\ell_2 m_2} a_{\ell_3 m_3} \right> = \mathcal{G}_{\ell_1 \ell_2 \ell_3}^{m_1 m_2 m_3} b_{\ell_1 \ell_2 \ell_3},
    \label{eq:gaunt_b}
\end{align}
with the Gaunt integral
\begin{align}
    &\mathcal{G}_{\ell_1 \ell_2 \ell_3}^{m_1 m_2 m_3} \equiv \int d\Omega\, Y_{\ell_1 m_1}(\hat{n}) Y_{\ell_2 m_2}(\hat{n}) Y_{\ell_3 m_3}(\hat{n}) \nonumber\\
    &= \mathcal{C}_{\ell_1 \ell_2 \ell_3} \six{\ell_1}{\ell_2}{\ell_3}{0}{0}{0}
    \six{\ell_1}{\ell_2}{\ell_3}{m_1}{m_2}{m_3},
    \label{eq:gaunt}
\end{align}
where $\int d\Omega$ represents integration over $\hat{n}$. Reading Eq. \ref{eq:gaunt_b} and Eq. \ref{eq:gaunt} together highlights two symmetries. First, the 3-$j$ symbol with the $m_i$, as previously discussed, enforces rotation invariance about the observer. Second, the 3-$j$ symbol with zeros enforces that $\ell_1 + \ell_2 + \ell_3$ is even; this means the CMB bispectrum is parity-even. These symmetries are reviewed in \cite{hu_01} Secs. IIA and B.

The CMB trispectrum $Q_{\ell_1 \ell_2 (\ell_{12})}^{\ell_3 \ell_4}$ \cite{hu_01} is defined by
\begin{align}
    &\left<a_{\ell_1 m_1} a_{\ell_2 m_2} a_{\ell_3 m_3} a_{\ell_4 m_4} \right> = \sum_{\ell_{12}, m_{12}} 
    \six{\ell_1}{\ell_2}{\ell_{12}}{m_1}{m_2}{-m_{12}}\nonumber\\
    &\qquad \times\six{\ell_{12}}{\ell_3}{\ell_4}{m_{12}}{m_3}{m_4}
    Q_{\ell_1 \ell_2 (\ell_{12})}^{\ell_3 \ell_4},
\end{align}
where relative to \cite{hu_01} we have switched the upper and lower indices of $Q$, rewritten $LM \to \ell_{12} m_{12}$ to highlight the coupling scheme used, and subscripted $\ell_{12}$ where \cite{hu_01} would have it appear like an argument of $Q$. As discussed in \cite{hu_01}, there are several choices for how to couple the ``external'' $\ell_i m_i$. These choices make no physical difference and are related by angular momentum recoupling with Wigner 6-$j$ symbols (discussion below Eq. 16 and in Appendix A of \cite{hu_01}). Here we adopt the coupling scheme of \cite{cahn_iso}, which matches that used in the body of \cite{hu_01}. We refer to $\ell_{12}$ as an ``intermediate'' and parentheses will distinguish intermediates throughout this work. 

We also note that the ``intermediate'' here has the same role as that appearing in the isotropic basis functions of 4 arguments of \cite{cahn_iso} (\cite{gen} gives the generating functions for this basis as well as further discussion of the role of intermediates). In \cite{cahn_iso}, the application of the basis functions is to 3D clustering, so each direction vector also has a length associated with it. The coupling scheme is then uniquely fixed by ordering the $\ell_i$ by the lengths corresponding to them, such that the $\ell_i$ corresponding to the smallest two lengths are coupled into the intermediate. For a given set of $\ell_i$ and lengths, any other coupling scheme would correspond to a different ordering of the lengths and can be converted to the fiducial one by a re-ordering operator defined in \cite{cahn_iso}. This operator is equivalent to the transformations between different coupling schemes in the CMB context in \cite{hu_01} (discussion above that work's Eq. 16).

\subsection{Smith-Zaldarriaga Algorithm}
It has for some years been of interest to produce a CMB map that has a given pair (power spectrum) and triplet (bispectrum) correlation function. The problem of realizing a given power spectrum is simple; one just draws $a_{\ell m}$ as an appropriate Gaussian Random Field (GRF). This problem was also solved in 3D, as appropriate for large-scale structure, with the well-known constrained realization algorithm of \cite{hr}. For the CMB bispectrum \cite{sz} gives an algorithm, which we recapitulate below.

\cite{sz} defines the functional
\begin{align}
    \label{eq:Tfunc}
    &T[\vec{a}] \equiv \frac{1}{6} \sum_{\ell_i m_i} B_{\ell_1 \ell_2 \ell_3}
    \six{\ell_1}{\ell_2}{\ell_3}{m_1}{m_2}{m_3} \nonumber\\
    &\qquad\qquad\qquad \times \ath
\end{align}
(their Eq. 55), where $\vec{a} = \{ a_{\ell_i m_i} \}$ is shorthand for the set of harmonic coefficients appearing on the righthand side, and $i = 1, 2, 3$. Relative to \cite{sz} we have replaced their $a$ with $\vec{a}$, as one can consider their set $a$ to be a vector in the space of $a_{\ell m}$. This notation emphasizes the involvement of multiple $a_{\ell m}$. We note that since the sum is over all $\ell_i m_i$, a particular given $\ell m$ may appear three times in the sum; we may have $\ell_1 m_1 = \ell m$, $\ell_2 m_2 = \ell m$, or $\ell_3 m_3 = \ell m$.

\cite{sz} also defines the operator 
\begin{align}
    &\nabla_{\ell m} T[\vec{a}] \equiv \frac{\partial T[\vec{a}]}{\partial a_{\ell m}^*} = \frac{1}{2}\sum_{\ell_i m_i} B_{\ell \ell_2 \ell_3}
    \six{\ell}{\ell_2}{\ell_3}{m}{m_2}{m_3}\nonumber\\
    &\qquad\qquad\qquad\qquad\qquad \qquad \times a_{\ell_2 m_2}^* a_{\ell_3 m_3}^*    
    \label{eq:nTdef}
\end{align} 
(their Eq. 56). The factor of $1/2$ comes from accounting for the fact that in the sum in Eq. \ref{eq:Tfunc}, a particular $\ell m$ can appear three times, and the derivative with respect to this $\ell m$ will pick up each of these, yielding $3 \times 1/6 = 1/2$. Eq. 56 of \cite{sz} seems to have a missing $\sum_{\ell_i m_i}$ on its righthand side, but this error does not affect that work's Eq. 84 because the expectation values there ($< a_{\ell_2' m_2'}^* a_{\ell_2 m_2} >$ and $< a_{\ell_3' m_3'}^* a_{\ell_3 m_3}>)$ would eliminate the sum in any case. We can be sure this sum is missing from the righthand side because without it, the number of free indices does not match between the lefthand and righthand sides. We also note that the derivative's being with respect to the conjugated $a_{\ell m}$ is the reason that the $a_{\ell_i m_i}$ on the right-hand side become conjugated.

\cite{sz} then shows that if one takes a GRF $a_{\ell m}$ and perturbs it by the gradient above (evaluated at the inverse power spectrum acting on the $a_{\ell m}$), at leading order in the perturbation, the resulting bispectrum will be $B_{\ell_1 \ell_2 \ell_3}$. Writing the perturbed $a_{\ell m}$ as
\begin{align}
    a_{\ell m}' \equiv a_{\ell m} + \frac{1}{3} \nabla_{\ell m} T [\vec{c}\,],
\end{align}
with 
\begin{align}
    \vec{c} \equiv \left\{C^{-1}_{\ell_i'} a_{\ell_i' m_i'} \right\},
\end{align}
with $i = 1, 2, 3$, we may compute 
\begin{align}
    \label{eq:aellmp}
    &\left< a_{\ell_1 m_1}' a_{\ell_2 m_2}' a_{\ell_3 m_3}' \right> = 
    \left< a_{\ell_1 m_1} a_{\ell_2 m_2} \frac{1}{3} \nabla_{\ell_3 m_3} T[ \vec{c}\,] \right > \nonumber\\
    &\qquad \qquad \qquad \qquad \qquad \qquad + 5 \; {\rm perms.}
\end{align}
Terms involving three unperturbed $a_{\ell m}$ vanish because $a_{\ell m}$ is a GRF and so has zero bispectrum. We have dropped terms involving two or three $\nabla_{\ell m}$ as being sub-leading, as these terms contain more $a_{\ell m}$ than the leading terms and $a_{\ell m}$ is presumed to be small compared to unity. The first term, and the five permutations, simply account for all orderings of $\ell_i m_i$

 The five permutations are the five other possible orderings of $\ell_i m_i$; including the first term, we w. 

Applying the definition Eq. \ref{eq:nTdef}, the first term of Eq. \ref{eq:aellmp} becomes
\begin{align}
\label{eq:interm_exp}
    &\left< a_{\ell_1 m_1} a_{\ell_2 m_2} \,\frac{1}{3} \nabla_{\ell_3 m_3} T[ \vec{c}\,] \right > = \nonumber\\
    &\qquad \frac{1}{6} \sum_{\ell_i'} B_{\ell_1 \ell_2' \ell_3'}
    \six{\ell_1}{\ell_2'}{\ell_3'}{m_1}{m_2'}{m_3'}\\
    &\qquad \times \left<  C_{\ell_2'}^{-1} a_{\ell_2 m_2} a_{\ell_2' m_2'}^* C_{\ell_3'}^{-1} a_{\ell_3 m_3} a_{\ell_3' m_3'}^* \right >.\nonumber
\end{align}
The additional factor of $1/2$ on the right-hand side came from the definition of $\nabla_{\ell m}$ in Eq. \ref{eq:nTdef}. Since $a_{\ell m}$ is a GRF, the expectation value may be pulled apart into two pairwise expectations:
\begin{align}
    &\left<  C_{\ell_2'}^{-1} a_{\ell_2 m_2} a_{\ell_2' m_2'}^* C_{\ell_3'}^{-1} a_{\ell_3 m_3} a_{\ell_3' m_3'}^* \right >
    = C_{\ell_2'}^{-1} \left<  a_{\ell_2 m_2} a_{\ell_2' m_2'}^* \right > \nonumber\\
    & \times C_{\ell_3'}^{-1} \left< a_{\ell_3 m_3} a_{\ell_3' m_3'}^* \right >
    = \delta^{\rm K}_{\ell_2 \ell_2'} \delta^{\rm K}_{m_2 m_2'}
    \delta^{\rm K}_{\ell_3 \ell_3'} \delta^{\rm K}_{m_3 m_3'},
\end{align}
with $\delta^{\rm K}$ the Kronecker delta, unity if its subscripts are equal and zero otherwise. Inserting this result in Eq. \ref{eq:interm_exp} we have
\begin{align}
    &\left< a_{\ell_1 m_1} a_{\ell_2 m_2} \frac{1}{3} \nabla_{\ell_3 m_3} T[ \vec{c}\,] \right > = \nonumber\\
    &\qquad \frac{1}{6} B_{\ell_1 \ell_2 \ell_3}
    \six{\ell_1}{\ell_2}{\ell_3}{m_1}{m_2}{m_3}.
\end{align}
The same process may be applied to the other five permutations in Eq. \ref{eq:aellmp}, so we obtain
\begin{align}
    &\left< a_{\ell_1 m_1}' a_{\ell_2 m_2}' a_{\ell_3 m_3}' \right> = B_{\ell_1 \ell_2 \ell_3}
    \six{\ell_1}{\ell_2}{\ell_3}{m_1}{m_2}{m_3},
\end{align}
\textit{i.e.} Eq. 84 of \cite{sz}. Recalling that the CMB power spectrum $C_{\ell}$ used here was arbitrary (and cancelled out of the CMB bispectrum obtained above), we see that this algorithm enables realizing a set of CMB $a_{\ell m}$ with any given power spectrum and bispectrum. 

\subsection{Towards a Generalization to 3D LSS}
We now seek to generalize this algorithm to 3D LSS. Let us notice that the CMB bispectrum actually involves four points: the observer, and the three points on the CMB spherical shell. This geometry is exactly the same as that used in the spherical-harmonic decomposition NPCF algorithms \cite{3pt_alg, 3pt_ft, encore, sarab}, but in that case, the ``primary'' galaxy is set to be at the origin and the ``secondaries'' are on spherical shells with different radii. In particular, the geometry for the CMB bispectrum is the same as that for the 4PCF around a given primary.

Let us briefly review the mathematical formulation pertinent to the discussion above. The NPCF algorithms of the series of papers noted above expand the density fluctuation field on each shell (designated by $r$) about a primary galaxy (at $\vec{x}$) as 
\begin{align}
    \delta(r, \hat{r}; \vec{x}) = \sum_{\ell m} a_{\ell m}(r; \vec{x}) Y_{\ell m}(\hat{r}).
    \label{eq:my_delta}
\end{align}
The NPCFs are then expanded in the isotropic basis of \cite{cahn_iso} (a generating function for, and further properties of, this basis are described in \cite{gen}). In particular, the 4PCF $\zeta$ is
\begin{align}
    \label{eqn:expansion}
    &\zeta(r_1, r_2, r_3; \hat{r}_1, \hat{r}_2, \hat{r}_3) \nonumber\\
    &\qquad = \left< \hat{\zeta}(r_1, r_2, r_3; \hat{r}_1, \hat{r}_2, \hat{r}_3; \vec{x}) \right>_{\vec{x}} \\
    &\qquad  = \left<  \delta(\vec{x}) \delta(r_1, \hat{r}_1; \vec{x})  \delta(r_2, \hat{r}_2; \vec{x})  \delta(r_3, \hat{r}_3; \vec{x})  \right>_{\vec{x}} \nonumber\\
    &\qquad \equiv \sum_{\ell_i} \zeta_{\ell_1 \ell_2 \ell_3}(r_1, r_2, r_3) \mathcal{P}_{\ell_1 \ell_2 \ell_3}(\hat{r}_1, \hat{r}_2, \hat{r}_3).\nonumber
\end{align}
Hat denotes the estimate of $\zeta$ about a specific primary galaxy at $\vec{x}$, and $\left<\cdots \right>_{\vec{x}}$ the average over all primaries. This last corresponds to averaging over translations, on the cosmological assumption of homogeneity. In the fourth line, we have performed the expansion into the isotropic basis $\mathcal{P}_{\ell_1 \ell_2 \ell_3}$ \cite{cahn_iso}, and this line may be taken to define the expansion coefficients $\zeta_{\ell_1 \ell_2 \ell_3}$. These are
\begin{align}
    \label{eq:zeta_ell}
    &\zeta_{\ell_1 \ell_2 \ell_3}(r_1, r_2, r_3) = 
    \left< \hat{\zeta}_{\ell_1 \ell_2 \ell_3}(r_1, r_2, r_3;\vec{x}) \right>_{\vec{x}}\nonumber\\
    &= \sum_{m_i} \six{\ell_1}{\ell_2}{\ell_3}{m_1}{m_2}{m_3} \nonumber\\
 &\qquad \times \left< a_{\ell_1 m_1}(r_1; \vec{x}) a_{\ell_2 m_2}(r_2; \vec{x}) a_{\ell_3 m_3}(r_3; \vec{x})\right>_{\vec{x}}.
\end{align}
To obtain the expansion coefficients, we first used orthogonality on Eq. \ref{eqn:expansion} and then applied Eq. 11 of \cite{cahn_iso} to convert $\mathcal{P}^*$ to $\mathcal{P}$; this is why the $a_{\ell m}$ are not conjugated here, and the phase in the definition of the 3-argument isotropic functions (\cite{cahn_iso} Eqs. 4 and 11) cancels. Finally, we used Eq. \ref{eq:my_delta} for each density to obtain an expression in terms of the harmonic coefficients of each density, the $a_{\ell m}$.

The similarity of the last line of Eq. \ref{eq:zeta_ell} to the CMB bispectrum is notable. It leads us to ask if the bispectrum algorithm of \cite{sz}, if generalized so that each $a_{\ell m}$ may lie on a different spherical shell, could enable realizing an arbitrary 4PCF about a given primary.
\section{\label{sec:4pcf} Realizing a Given 4PCF}
We first make a comment about permutations. For the CMB, there is no way to distinguish the $\ell_i m_i$, so we end up with the sum over permutations of them. Once we are working on spherical shells of different radii $r_i$, we may let the $r_i$ give a unique ordering to the $\ell_i$. Indeed, this is needed to uniquely define the coupling scheme used for the isotropic basis functions in which the harmonic-based algorithms measure NPCFs, as detailed in \cite{cahn_iso} and briefly reviewed in the last paragraph of Sec. \ref{subsec:cmb_corr} here. 

For the 4PCF, let us choose the convention that we consider only configurations of three galaxies around a given primary where the relative distances $r_i$ satisfy $r_1 \leq r_2 \leq r_3$. This then gives unique, physically observable meaning to $\ell_1, \ell_2$, \textit{etc.} $\ell_1$ is the $\ell$ associated with the harmonic decomposition of the pattern of secondaries on the smallest shell of a set of three, $\ell_2$ of the second-smallest, \textit{etc.} This convention eliminates the need for permutations over the $\ell_i$, rendering the LSS problem actually simpler than the CMB problem.

In typical analyses of the 4PCF, such as \cite{hou_odd, phil_odd, phil_even}, one avoids configurations where $r_i = r_j$, $i \neq j$ as this can allow two galaxies to be arbitrarily close to each other. If they are arbitrarily close, they may then be well into the regime of non-linear structure formation, where both any possible signal, and the covariance matrix, become unreliable to model due both to non-linearity and possible baryonic physics effects \cite{hou_odd, hou_cov, phil_even}. We note that \cite{will} gives a model of the tree-level redshift-space ``connected'' (even-parity, due to non-linear structure formation) 4PCF, which might be one desirable choice to impose using the algorithm of the present work. 

{\bf Thus, in all that follows in this work, we assume that the ordering of $r_i$ is strict, \textit{i.e.} $r_1  < r_2 < r_3$ (and $r_3 < r_4$ when we move to the 5PCF in Sec. \ref{sec:npcf})}.

By analogy with Sec. \ref{sec:cmb}, let us define 
\begin{align}
    &\mathcal{T}^{(4)}[\vec{a}] \equiv \sum_{\ell_i' m_i', r_i'} \zeta_{\ell_1' \ell_2' \ell_3'}(r_1', r_2', r_3') 
    \six{\ell_1'}{\ell_2'}{\ell_3'}{m_1'}{m_2'}{m_3'} \nonumber\\
    &\qquad \qquad \times \athrprime.
    \label{eq:calT}
\end{align}
We note that any particular given value of $\ell m$ may appear in three different positions: $\ell_1' m_1' = \ell m$, $\ell_2' m_2' = \ell m$, and $\ell_3' m_3' = \ell m$, just as in the CMB algorithm. However, the ordering by $r_i'$ means there is no need to consider permutations.

We also define
\begin{align}
    \label{eq:my_nabla}
    &\nabla_{\ell m, r} \mathcal{T}^{(4)} [\vec{a}] \equiv \frac{\partial \mathcal{T}^{(4)} [\vec{a}]}{\partial (a_{\ell m}^*(r))} = \sum_{\ell_i' m_i', r_i'} 
    \zeta_{\ell \ell_2' \ell_3'}(r, r_2', r_3') \nonumber\\
    &\qquad \qquad \times \six{\ell}{\ell_2'}{\ell_3'}{m}{m_2'}{m_3'} a_{\ell_2' m_2'}^*(r_2') a_{\ell_3' m_3'}^*(r_3').
\end{align}
The derivative is intended to be for a specific $\ell m$ and shell $r$. The conjugate in the derivative is what converts the $a_{\ell m}$ on the right-hand side into their conjugates, relative to Eq. \ref{eq:calT}.

Let us now consider perturbing the $a_{\ell m}$ on each shell (denoted by $r$) about a primary galaxy at $\vec{x}$ as
\begin{align}
    a_{\ell m}'(r; \vec{x}) = a_{\ell m}(r; \vec{x}) + \frac{1}{3} \nabla_{\ell m, r} \mathcal{T}^{(4)}[ \vec{c}\,^{(4)}], 
    \label{eq:my_pert}
\end{align}
where we have generalized the $\vec{c}$ of Sec. \ref{sec:cmb} to 
\begin{align}
   \vec{c}\,^{(4)} = \left\{C_{\ell_i'}^{-1}(r_i') a_{\ell_i' m_i'}(r_i'; \vec{x}) \right\}
   \label{eq:vecc}
\end{align}
and $i = 1, 2, 3$. Again, by assumption, $a_{\ell' m'}(r_i')$ is a GRF with power spectrum on that shell $C_{\ell_i'}(r_i')$. Importantly, the perturbation on a given shell depends only on that shell's $r_i'$ and $\ell' m'$. Going back to the definition Eq. \ref{eq:my_nabla} we see the perturbation is a sum over all $r_2', r_3'$ and $\ell_2' m_2', \ell_3' m_3'$, \textit{i.e.} all the shells not picked out by the derivative.

We now compute the leading-order 4PCF coefficient about the primary, that is, Eq. \ref{eq:zeta_ell} prior to averaging over $\vec{x}$. Let us first focus on the last line of Eq. \ref{eq:zeta_ell}, and write it prior to averaging over $\vec{x}$. We have, dropping terms with three unperturbed $a_{\ell m}$ (as they will vanish after averaging), and dropping sub-leading terms, that
\begin{align}
    &a_{\ell_1 m_1}'(r_1; \vec{x})a_{\ell_2 m_2}'(r_2; \vec{x})a_{\ell_3 m_3}'(r_3; \vec{x}) = \nonumber\\
    & a_{\ell_1 m_1}(r_1; \vec{x}) a_{\ell_2 m_2}(r_2; \vec{x}) 
    \frac{1}{3} \nabla_{\ell_3 m_3, r_3} \mathcal{T}^{(4)}[ \vec{c}\,^{(4)}] \nonumber\\
    & \qquad + 2\;{\rm perms.}
\end{align}
Using the definition Eq. \ref{eq:my_nabla} we find
\begin{align}
    \label{eq:4pcf_interm}
    &a_{\ell_1 m_1}'(r_1; \vec{x})a_{\ell_2 m_2}'(r_2; \vec{x})a_{\ell_3 m_3}'(r_3; \vec{x}) = \nonumber\\
    & a_{\ell_1 m_1}(r_1; \vec{x}) a_{\ell_2 m_2}(r_2; \vec{x})\,
    \frac{1}{3} \sum_{\ell_i' m_i', r_i'} \zeta_{\ell_1' \ell_2' \ell_3} (r_1', r_2', r_3)\nonumber\\
     &\qquad \times \six{\ell_1'}{\ell_2'}{\ell_3}{m_1'}{m_2'}{m_3} C_{\ell_1'}^{-1}(r_1') C_{\ell_2'}^{-1}(r_2')\nonumber\\
     &\qquad \times a_{\ell_1' m_1'}^*(r_1'; \vec{x})
     a_{\ell_2' m_2'}^*(r_2'; \vec{x})+ 2\;{\rm perms.}
\end{align}
We emphasize that the sum in the second line is not over $\ell_3$ or $r_3$; taking the derivative has picked just one term in these variables out from the sum in Eq. \ref{eq:calT} (and we recall Eq. \ref{eq:my_nabla}). 

Now, inserting Eq. \ref{eq:4pcf_interm} in Eq. \ref{eq:zeta_ell} and taking the expectation value, we find
\begin{align}
    \label{eq:four_exp}
    &\left< a_{\ell_1 m_1}'(r_1; \vec{x})a_{\ell_2 m_2}'(r_2; \vec{x})a_{\ell_3 m_3}'(r_3; \vec{x}) \right>_{\vec{x}} = \nonumber\\
    &\qquad \frac{1}{3} \sum_{\ell_i' m_i', r_i'} \zeta_{\ell_1' \ell_2' \ell_3} (r_1', r_2', r_3)\nonumber\\
     &\qquad \times \six{\ell_1'}{\ell_2'}{\ell_3}{m_1'}{m_2'}{m_3} C_{\ell_1'}^{-1}(r_1') C_{\ell_2'}^{-1}(r_2')\nonumber\\
     &\qquad \times \left< a_{\ell_1 m_1}(r_1; \vec{x}) a_{\ell_2 m_2}(r_2; \vec{x})  a_{\ell_1' m_1'}^*(r_1'; \vec{x})
     a_{\ell_2' m_2'}^*(r_2'; \vec{x})\right>_{\vec{x}} \nonumber\\
     & \qquad \qquad \qquad + 2\;{\rm perms.}
\end{align}
Now, the $a_{\ell m}$ are, by construction, not correlated across shells, so we must have $r_1 = r_1', r_2 = r_2'$. Further, since on each shell, the field is a GRF, we have Kronecker deltas setting $\ell_1 m_1 =  \ell_1' m_1'$ and $\ell_2 m_2 = \ell_2' m_2'$ as in the CMB case. Thus the sum over $\ell_i' m_i', r_i'$ drops out and when the conditions above are satisfied, we simply get $C_{\ell_1}(r_1) C_{\ell_2}(r_2)$ for the expectation value in the fourth line above. We thus have
\begin{align}
    &\left< a_{\ell_1 m_1}'(r_1; \vec{x})a_{\ell_2 m_2}'(r_2; \vec{x})a_{\ell_3 m_3}'(r_3; \vec{x}) \right>_{\vec{x}} = \nonumber\\
    & \qquad  \frac{1}{3}\, \zeta_{\ell_1 \ell_2 \ell_3} (r_1, r_2, r_3) \six{\ell_1}{\ell_2}{\ell_3}{m_1}{m_2}{m_3} \nonumber\\
     & \qquad\qquad + 2\;{\rm perms.}\nonumber\\
     &\qquad = \zeta_{\ell_1 \ell_2 \ell_3} (r_1, r_2, r_3) \six{\ell_1}{\ell_2}{\ell_3}{m_1}{m_2}{m_3}.
\end{align}
Inserting this result for the average of the three $a_{\ell m}'$ in Eq. \ref{eq:zeta_ell}, we have
\begin{align}
    \label{eq:zeta_ell_res}
   &\zeta_{\ell_1 \ell_2 \ell_3}(r_1, r_2, r_3) \sum_{m_i} \six{\ell_1}{\ell_2}{\ell_3}{m_1}{m_2}{m_3}  \six{\ell_1}{\ell_2}{\ell_3}{m_1}{m_2}{m_3}\nonumber\\
   & = \zeta_{\ell_1 \ell_2 \ell_3} (r_1, r_2, r_3) 
\end{align}
using the orthogonality identity 34.3.18 in \cite{nist} for the 3-$j$ symbols. 

Thus, we see that producing a GRF $a_{\ell m}(r)$ on each shell (independent on each shell), and perturbing on each shell as in Eq. \ref{eq:my_pert}, will produce the desired 4PCF, with coefficients $\zeta_{\ell_1 \ell_2 \ell_3}$ when expanded in the isotropic basis of \cite{cahn_iso}. We emphasize that the perturbation on a given shell $r$ depends only on that shell's $r$; it is a sum over all the other shells' $r_i$. Thus, it can give the correct desired 4PCF at all combinations of $r_i$ shells; it is not just tuned to give the right 4PCF at only one combination.

Finally, we emphasize that no 3-$j$ symbol with all zero $m$ ever appeared; thus there is no restriction on the sum of the $\ell_i$, and so we can realize a parity-odd 4PCF of the kind considered in \cite{cahn_short,hou_odd,phil_odd}. This is valuable for the search for parity-violation in the 4PCF (parity-violating 4PCF coefficients have odd $\ell_1 + \ell_2 + \ell_3$ \cite{cahn_iso}, which would not be allowed if a 3-$j$ symbol with all zero $m_i$ appeared).

\section{\label{sec:3pcf} Realizing a Given 3PCF}
The harmonic-based 3PCF algorithm \cite{3pt_alg,3pt_ft,sarab} essentially sits on a primary galaxy at $\vec{x}$ and computes angular cross power spectra of two different spherical shells, $r_1$, $r_2$. Thus, it is exactly like the CMB angular power spectrum problem with the observer being analogous to the primary galaxy, and if we generalize to two shells rather than just one. This provokes wondering whether simply promoting the unperturbed $a_{\ell m}$ of Sec. \ref{sec:4pcf} to have a non-vanishing power spectrum across shells would enable realizing a desired 3PCF, \textit{i.e.} allowing $<a_{\ell m} (r_1) a_{\ell m}^*(r_2)> \neq 0$. 

However, this will cause the 4PCF algorithm of Sec. \ref{sec:4pcf} to break down, as there we needed to assume that the $a_{\ell m}$ are independent on each shell so that only the desired $r_i$ shells get picked out from the sum in Eq. \ref{eq:my_nabla}. We might consider removing this sum from the definition \ref{eq:my_nabla} and simply defining that quantity at a desired set of three shells, but then our 4PCF algorithm would only reproduce the desired 4PCF at one set of shells at a time. So we need a different strategy.

The most obvious option is simply to add an additional field to the $a_{\ell m}$ as a perturbation, as we did in the 4PCF case. This turns out to have the desired effect.

When expanded in the isotropic basis of two-argument functions $\mathcal{P}_{\ell \ell}(\hat{r}_1, \hat{r}_2)$ \cite{cahn_iso} (or, up to a scaling and phase, the Legendre basis \cite{3pt_alg, 3pt_ft}; the relationship is in \cite{cahn_iso} Eq. 3), the 3PCF is
\begin{align}
\zeta(r_1, r_2; \hat{r}_1 \cdot \hat{r}_2)  = \sum_{\ell} \zeta_{\ell}(r_1, r_2) \mathcal{P}_{\ell \ell}(\hat{r}_1, \hat{r}_2).  
\label{eq:3pcf_sum}
\end{align}
The expansion coefficient is
\begin{align}
    \label{eq:zeta3_ell}
    &\zeta_{\ell}(r_1, r_2) = 
    \left< \hat{\zeta}_{\ell }(r_1, r_2;\vec{x}) \right>_{\vec{x}}\nonumber\\
    &= \frac{(-1)^{\ell}}{\sqrt{2 \ell + 1}} \sum_{m}  \left< a_{\ell m}(r_1; \vec{x}) a_{\ell m}^*(r_2; \vec{x})\right>_{\vec{x}}.
\end{align}
We define
\begin{align}
    \mathcal{T}^{(3)} [\vec{a}] \equiv  \sum_{r_i', \ell' m'} \zeta_{\ell'} (r_1', r_2') a_{\ell' m'}(r_1'; \vec{x}) a_{\ell' m'}(r_2'; \vec{x})
\end{align}
and 
\begin{align}
\label{eq:3pcf-nabla}
    \nabla_{\ell m, r} \mathcal{T}^{(3)}  [\vec{a}] \equiv \frac{\partial \mathcal{T}^{(3)}}{a_{\ell m}^*(r)} = \sum_{r_i'} \zeta_{\ell}(r, r_2') a_{\ell m}^*(r_2'; \vec{x}).
\end{align}
We then perturb the $a_{\ell m}$ as
\begin{align}
    a_{\ell m}'(r; \vec{x}) = a_{\ell m}(r; \vec{x}) + \frac{1}{2} \nabla_{\ell m, r} \mathcal{T}^{(3)} [\vec{c}\,^{(3)}]
\end{align}
with $\vec{c}\,^{(3)}$ analogous to that in Eq. \ref{eq:vecc}: 
\begin{align}
    \vec{c}\,^{(3)} \equiv \left\{ C_{\ell'}^{-1}(r_i') a_{\ell_i' m_i'}(r_i'; \vec{x}) \right\}
\end{align}
with $i = 1, 2$. We then have that
\begin{align}
    \label{eq:3pcf_prod}
    &\left< a_{\ell m}'(r_1; \vec{x}) (a_{\ell m}'(r_2; \vec{x}))^* \right> = \frac{1}{2} \sum_{r_i'} \zeta_{\ell}(r_1', r_2)  \, C_{\ell'}^{-1}(r_1')\nonumber \\
    & \qquad\qquad \times \left< a_{\ell m}(r_1; \vec{x}) ( a_{\ell' m'}(r_1'; \vec{x}))^*  \right>  + 1 {\;\rm perm.}\nonumber\\
    & = \zeta_{\ell}(r_1, r_2).
\end{align}
The conjugate on the $a_{\ell' m'}$ on the right-hand side comes from the derivative with respect to the conjugate required by the definition Eq. \ref{eq:3pcf-nabla}, while the last line follows by noticing that the only non-vanishing expectation of the $a_{\ell m}$ will force $\ell' = \ell$ and $r_1' = r_1$. 

Now, if we wish to realize both a given 3PCF and a given 4PCF at the same time, we would perturb the $a_{\ell m}$ as 
\begin{align}
    &a_{\ell m}'(r; \vec{x}) = a_{\ell m}(r; \vec{x}) + \frac{1}{2} \nabla_{\ell m, r} \mathcal{T}^{(3)} [\vec{c}\,^{(3)}]\nonumber\\
    & \qquad \qquad \qquad +  \frac{1}{3} \nabla_{\ell m, r} \mathcal{T}^{(4)} [\vec{c}\,^{(4)}].
\end{align}
If we form the pair correlation of $a_{\ell m}'$ required for the 3PCF, any terms due to the 4PCF will vanish to leading order because they will involve one $a_{\ell m}$ and one $\nabla_{\ell m, r} \mathcal{T}^{(4)}$, which latter contains two $a_{\ell m}$. Thus we have an odd correlator, which vanishes for a GRF. For a similar reason, the perturbation required to produce the 3PCF (\textit{i.e.} $\nabla_{\ell m, r} \mathcal{T}^{(3)}$) will vanish from the 4PCF; it contains one $a_{\ell m}$, and the other two $a_{\ell'm'}$ required when we form the 4PCF will also contain one each, so we obtain an odd correlator, which vanishes.

\section{\label{sec:factor} Factorizability and Fast Realization}
\cite{sz} presents a scheme showing how to realize the needed $\nabla_{\ell m} T$ field fast if the desired CMB bispectrum is factorizable, and also develops a scheme to approximately factorize non-factorizable templates by writing them as an integral over a dummy variable and then using a discretization of the integral to get a sum of factored terms. This relies on the fact that the integrand of this integral is itself factorizable. Here, we show that the tree-level 3PCF template, and that certain recently-used parity-odd 4PCF templates, can be treated in the same way. 
\subsection{\label{subsec:3pcf_fac} 3PCF}

\subsubsection{\label{subsubsec:3pcf_red} Reduction to Integrals with Factorizable Integrands}
The 3PCF model of \cite{rsd_3pcf}, which is the tree-level standard perturbation theory prediction in redshift space, has initially non-factorizable behavior. We use a trick similar to that of \cite{sz} to address this. \cite{sz} writes any non-factorizable CMB bispectrum as an integral over a factorizable integrand. \cite{sz} then notes that if the integral is performed numerically with quadrature, that integral becomes a sum over the integrand evaluated at the quadrature points, and each term in that sum is factorizable. This then gives the form required for their algorithm, where the CMB bispectrum is represented as a sum of factorizable pieces.

We observe that, if the post-cyclic 3PCF is written in the form Eq. \ref{eq:3pcf_sum}, each of the coefficients $\zeta_{\ell}(r_1, r_2)$ will be a sum of terms derived from projecting the pre-cyclic 3PCF at $\vec{r}_2, \vec{r}_3$ and $\vec{r}_3, \vec{r}_1$ back onto the isotropic basis at $\hat{r}_1, \hat{r}_2$ (\textit{e.g.} discussed in Sec. 3 of \cite{rsd_3pcf}). Each such term will look like \footnote{We neglect terms that stem from a $k_3$-dependence in the pre-cyclic bispectrum. These terms, written using functions $\kappa_{\ell}(r_1, r_2)$ defined in \cite{rsd_3pcf}, are not factorizable even at the pre-cyclic level. However, they are suppressed by three orders of magnitude and so can safely be neglected; compare Figs. 1 and 2 of \cite{rsd_3pcf}. As discussed in \cite{rsd_3pcf}, cyclic summing and re-projection of these terms could be handled with the same approach as developed for the others in \cite{rsd_3pcf}.  Thus the derivation in the main text of the present work can be used on them too if desired.}
\begin{align}
    &\xi^{[L]}(r_1) \xi^{[L]}(r_3) \mathcal{P}_L(\hat{r}_1, \hat{r}_3) = \nonumber\\
    &\sum_{J} \zeta_{J}(r_1, r_2)  \mathcal{P}_J(\hat{r}_1, \hat{r}_2)
\end{align}
with coefficients
\begin{align}
    \label{eq:post_cyc_coeffs}
    &\zeta_{J}(r_1, r_2) = (-1)^{\ell} \frac{4\pi}{\sqrt{2\ell + 1}} \xi^{[L]}(r_1) \sum_{L_1} \mathcal{C}_{L J L_1}^2 \six{L}{J}{L_1}{0}{0}{0}^2\nonumber\\
    &\qquad \times \int \frac{k^2 dk}{2\pi^2}\; P(k) j_{L_1}(kr_1) j_{J}(kr_2),
\end{align}
as in \cite{rsd_3pcf} Eq. 30; this coefficient is specifically from the projection of one of the two cyclic terms (the one in $\vec{r}_1, \vec{r}_3$) at a pre-cyclic multipole $L$ onto $P_J(\hat{r}_1 \cdot \hat{r}_2)$; we have updated the formula of \cite{rsd_3pcf} Eq. 30 to reflect that we are projecting onto the isotropic basis here, not Legendre polynomials as in that work. We recall that $\mathcal{C}_{L J L_1}$ is defined in Eq. \ref{eq:calC}.  

If this integral is performed by numerical quadrature, then we will have a sum over the quadrature points in $k$, and at each such point, the integrand is factorizable. 

\subsubsection{\label{subsubsec:3pcf_num} Numerical Considerations for Choosing the Quadrature Points}
The spherical Bessel functions, at a given $r$, are band-limited functions of $k$, since they are simply the projection of a plane wave along a line of sight that is not necessarily perpendicular to the wave-fronts; thus their frequencies are set by that of the underlying plane wave. Put mathematically, at each $r$, they are just sums of cosines and sines of $kr$ weighted by rational functions of $kr$. Thus, given some $r_{\rm max}$ representing the maximal distance out to which we wish to measure our 3PCF, we may set an exact sampling scheme for them. Standard tricks like the Poisson summation formula may be employed to accelerate the convergence of the sum by which we numerically evaluate the integral. 

The power spectrum also should be considered when choosing quadrature points; it is a smooth envelope, with a small additional ``rider'' of BAO wiggles beginning at $k \simeq 0.01\; h/{\rm Mpc}$. Since for numerical integrations, $P(k)$ is always smoothed by a function such as $\exp[-k^2 \sigma^2]$ with $\sigma \simeq 1 \;{\rm Mpc}/h$ to avoid Gibbs' phenomenon ringing, $P(k)$ effectively has roughly compact support, further reducing the number of quadrature points needed. Often a logarithmic grid is employed. Typically of order a hundred or so points is sufficient. 

Conveniently, $r_{\rm max}$ we usually employ for 3PCF measurements (\textit{e.g.} \cite{se_bao, se_boss, moresco, 3pt_alg} is roughly twice the BAO scale (\textit{i.e.} $\sim 200\;{\rm Mpc}/h$), meaning that the frequency of the sBFs will at most be twice or so the BAO wiggles' frequency. Now, the two sBFs in Eq. \ref{eq:post_cyc_coeffs} can resonate (consider multiplying out the sines and cosine and reducing the products to sums using the double-angle formulae), giving another factor of two. Thus, a sampling four times as fine as that needed for the BAO should suffice.

Typically a numerical integration might need to range from $k_{\rm min} \simeq 10^{-4} \;h/{\rm Mpc}$ to $k_{\rm max} \sim 3\;h/{\rm Mpc}$, and sample logarithmically. The BAO in the transfer function has of order ten peaks from $0.01 \;h/{\rm Mpc}$ to $0.1 \;h/{\rm Mpc}$ (\cite{eh} Fig. 3); the power spectrum is the transfer function's square and so with resonance of the oscillations, can have twice that. We then need to sample at roughly four times the frequency of power spectrum peaks for the reasons in the previous paragraph. Thus, we should have roughly 80 points from $0.01 \;h/{\rm Mpc}$ to $0.1 \;h/{\rm Mpc}$. If we sample uniformly per dex, and have about 5 dex from $k_{\rm min}$ to $k_{\rm max}$, this implies 400 quadrature points. Thus, the integral required by Eq. \ref{eq:post_cyc_coeffs} should be well-approximated by a sum over 400 terms, each of which is factorizable. This is comparable in order of magnitude to the number of points used for the \textit{Planck} forecast in \cite{sz} (discussion following their Eq. 44), and likely can be improved upon significantly using \textit{e.g.} Poisson summation.

\subsection{\label{subsec:4pcf_fac} 4PCF}
Recently \cite{cabass} presented a search for several parity-violating inflationary trispectrum templates. For instance, their Eq. 11 presents one for a ghost condensate, with radial ($k$-space) coefficients in their Eq. 12. The angular part of their template can be decomposed into the three-argument isotropic basis of \cite{cahn_iso}, and so we are left in Fourier space with the radial part. This has the form of an integral over $\lambda$ that mixes factors involving $\lambda k_1, \lambda k_2, \lambda k_3$, and  $\lambda k_4$. $k_4$ is constrained as $|\vec{k_1} + \vec{k}_2 + \vec{k}_3|$. We may formally ``free'' $k_4$ by introducing a 3D Dirac delta function, rewriting as the FT of unity, and using this to write its radial coupling (between the $k_i$) as a quadruple-sBF integral over a dummy variable $r$, much as was done for a three-argument 3D delta function in \cite{3pt_alg} Eq. 65.

Then the integrand of Eq. 11 of \cite{cabass} is formally factorized in Fourier space. We may take the inverse FT to configuration space separately on each $k_i$, delaying the coupling integral over $r$ and the integral over $\lambda$ for after that. The integrand of this ``delayed'' double integral will be formally factorized in configuration space, and can be done with a double quadrature. Though it is more expensive than a single integral, we are aided by the fact that all the functions involved are known analytically in closed form, in contrast to the power spectrum entering the 3PCF model of Sec. \ref{subsec:3pcf_fac}. Thus an optimal quadrature scheme is easier to obtain. 

The same approach as above can be used on the cosmological collider template of \cite{cabass}, their Eq. 18. This template, in addition to $k_4$, depends on the magnitudes of pair-wise combinations of wave-vectors, \textit{e.g.} $k_{12}$, necessitating an additional delta function. We may rewrite each delta function as the FT of unity as in the first paragraph of this sub-section. There is no integral over $\lambda$; in Eq. 18 of \cite{cabass}. Once the two delta functions are expanded, the template is thus simply a product of factors in Fourier space. So we will again have only a double integral. Here, the template does include the primordial curvature power spectrum, in contrast to the earlier one, which only contains known analytic functions. However, since the primordial curvature power spectrum is very nearly a power law, with no oscillations, it does not impact the easiness of finding an optimal sampling scheme for the quadrature.

\section{\label{sec:2pcf} Including the 2PCF}
We now discuss how to include a desired 2PCF about the primary galaxy at $\vec{x}$. About a primary at $\vec{x}$, for a GRF on each shell $a_{\ell m}(r; \vec{x})$, the 2PCF between the primary and that shell will simply be
\begin{align}
    \xi_0(r;\vec{x}) = \delta(\vec{x}) a_{00}(r; \vec{x}),
\end{align}
where $\delta(\vec{x})$ is the density fluctuation associated with the primary and only $a_{00}$ contributes because the 2PCF is an isotropic average. 

We recall that the 4PCF is actually independent of the angular power spectrum $C_{\ell}(r)$ of the $a_{\ell m}(r)$. Thus we may give $a_{00}$ some desired ``global'' radial dependence without affecting the 4PCF, as the inverse power spectrum there will cancel it out. Thus, we may realize any desired 2PCF along with any 3PCF and 4PCF. We do note that there is an effect on the 3PCF monopole when $r_1 = r_2$. However, as discussed in Sec. \ref{sec:4pcf}, third paragraph, in typical 4PCF analyses, the limit where $r_i = r_j$ is avoided, and this is also true for the 3PCF \cite{se_bao, se_boss, rsd_3pcf}.


\section{\label{sec:npcf} Extending to NPCFs, $N>4$: 5PCF as an Example}
The fundamental requirements of the algorithm of this work are simply that the NPCFs be estimated around a primary galaxy by decomposing the density field on spherical shells in spherical harmonics (an approach first proposed for the 3D LSS 3PCF in \cite{3pt_alg}, building on the Legendre basis of \cite{szap}, and extended to use FTs in \cite{3pt_ft}). Thus, the algorithm trivially extends to NPCFs for $N > 4$. We note this is useful because \cite{encore} indeed can measure the 5PCF and 6PCF in the isotropic basis of \cite{cahn_iso}.

Here we write down the analog of $\mathcal{T}$ (Eq. \ref{eq:calT}) required for realizing a given 5PCF as well; going to $N>5$ is straightforward but the expressions are lengthy. 

First of all, we note that the 5PCF is expanded in the four-argument isotropic basis functions $\mathcal{P}_{\ell_1 \ell_2 (\ell_{12}) \ell_3 \ell_4}$ \cite{cahn_iso} as
\begin{align}
    &\zeta(r_1, r_2, r_3, r_4; \hat{r}_1, \hat{r}_2, \hat{r}_3, \hat{r}_4) = \\
    & \sum_{\ell_i, \ell_{12}}  \zeta_{\ell_1 \ell_2 (\ell_{12}) \ell_3 \ell_4}(r_1, r_2, r_3, r_4)\nonumber\\ &\qquad \times \mathcal{P}_{\ell_1 \ell_2 (\ell_{12}) \ell_3 \ell_4}(\hat{r}_1, \hat{r}_2, \hat{r}_3, \hat{r}_4).\nonumber
\end{align}
We have 
\begin{align}
    \label{eq:calT5}
    &\mathcal{T}^{(5)} [\vec{c}\,^{(5)}] \equiv \sum_{\ell_i' m_i', r_i'} \sum_{\ell_{12}', m_{12}'} 
    \six{\ell_1'}{\ell_2'}{\ell_{12}'}{m_1'}{m_2'}{-m_{12}'}\\
    &\qquad \times\six{\ell_{12}'}{\ell_3'}{\ell_4'}{m_{12}'}{m_3'}{m_4'}\zeta_{\ell_1' \ell_2' (\ell_{12}) \ell_3' \ell_4'}(r_1', r_2', r_3', r_4') \nonumber\\
    &\qquad \qquad \times \afrprime. \nonumber
\end{align}
$\zeta_{\ell_1' \ell_2' (\ell_{12}') \ell_3' \ell_4'}$ is the coefficient in the isotropic basis of the 5PCF we wish to impose. We also recall that $\ell_{12}'$ is an ``intermediate'' (discussion in Sec. \ref{sec:cmb} and in \cite{cahn_iso} and \cite{gen}). 

Defining $\nabla_{\ell m, r}$ analogously to Eq. \ref{eq:my_nabla}, we have
\begin{align}
    \label{eq:my_nabla_5}
    &\nabla_{\ell m, r} \mathcal{T}^{(5)} [\vec{c}\,^{(5)}] \equiv \frac{\partial \mathcal{T}^{(5)} [\vec{c}\,^{(5)} ]}{\partial (a_{\ell m}^*(r))} \nonumber\\
    &\qquad\qquad = \sum_{\ell_i' m_i', r_i'} \sum_{\ell_{12}', m_{12}'} 
    \six{\ell}{\ell_2'}{\ell_{12}'}{m}{m_2'}{-m_{12}'}\nonumber\\
    &\qquad\qquad \times\six{\ell_{12}'}{\ell_3'}{\ell_4'}{m_{12}'}{m_3'}{m_4'}\zeta_{\ell \ell_2' (\ell_{12}') \ell_3' \ell_4'}(r, r_2', r_3', r_4')\nonumber\\
    & \qquad\qquad \times a_{\ell_2' m_2'}^*(r_2') a_{\ell_3' m_3'}^*(r_3')a_{\ell_4' m_4'}^*(r_4').
\end{align}
The computation of the 5PCF about a primary proceeds analogously to that in Sec. \ref{sec:4pcf}. We note that at leading order, adding a term like this (with a pre-factor of $1/4$ to account for the four different times one factor of $\nabla_{\ell m, r}$ can enter the correlation) will not affect the lower-order NPCFs, with the argument being analogous to that of Sec. \ref{sec:3pcf}; we sketch this further beginning in the second paragraph below Eq. \ref{eq:5pcf_prod}.

We now outline the 5PCF computation. We write
\begin{align}
    \label{eq:my_pert_five}
    &a_{\ell m}'(r; \vec{x}) = a_{\ell m}(r; \vec{x}) \\
    &\qquad \qquad \qquad \qquad + \frac{1}{4} \nabla_{\ell m, r} \mathcal{T}^{(5)}[ \vec{c}\,^{(5)}],\nonumber  
\end{align}
with 
\begin{align}
    \vec{c}\,^{(5)} \equiv \left\{ C_{\ell_i'}^{-1} a_{\ell_i' m'_i}(r_i'; \vec{x}) \right\}
\end{align}
and $i  = 1, 2, 3, 4$.

The 5PCF is proportional to
\begin{align}
    \left< a_{\ell_1 m_1}'(r_1; \vec{x}) \cdots a_{\ell_4 m_4}'(r_4; \vec{x})  \right>.
    \label{eq:5pcf_prod}
\end{align}
At leading order, we consider only correlators with one $\nabla_{\ell m, r}$. The unperturbed $a_{\ell m}$ contributes a product of two pair correlations; however, this is non-vanishing only if two of the shells are equal, a limit we avoid in this work. We also note that the correct intermediate, $\ell_{12}$, will be picked out by the orthogonality of the 3-$j$ symbols that enter the four-argument basis function we are projecting our density onto with the 3-$j$ symbols that enter the perturbed $a_{\ell m}$. 

We now sketch the proofs that i) the 3PCF and 4PCF perturbations do not affect the 5PCF and ii) the 5PCF perturbation does not affect the 3PCF or 4PCF. 

i) When we compute Eq. \ref{eq:5pcf_prod}, the 3PCF perturbation (if included in $a_{\ell m}'$) will contribute a $\nabla_{\ell m}$ to a term that then has three unperturbed $a_{\ell m}$; this gives four overall, which may be non-vanishing for a GRF. However, at least two of the unperturbed $a_{\ell m}$ will be on different bins, and thus their expectation value will vanish. 

The 4PCF perturbation, if included in $a_{\ell m}'$, will result in a term with five $a_{\ell m}$ overall when we compute Eq. \ref{eq:5pcf_prod}; these vanish because it is an odd correlator of a GRF.

ii) If the 5PCF perturbation is included in $a_{\ell' m'}$ when we compute the 3PCF, its $\nabla_{\ell m, r}$ will contribute three $a_{\ell m}$, and the other part of Eq. \ref{eq:3pcf_prod} will contribute another. This will give an even correlator, but at least two of the shells will be different and so the expectation value will vanish. 

If included when the 4PCF is computed, the 5PCF perturbation will result in terms with overall five $a_{\ell m}$, which will vanish in a GRF.

\section{\label{sec:dpcf} Extending to Dimensions Greater than 3 and non-scalar fields}
\cite{phil_nd} shows how to extend the harmonic-based algorithm to arbitrary spatial dimension and arbitrary spatial geometry, via hyperspherical harmonics and generalizations of the Wigner 3-$j$ symbols. Since the algorithm presented in the present paper for realizing NPCFs requires only the same factorization properties as the 3D, flat space version, extension of our approach for realizing higher-D NPCFs is straightforward. 

Indeed, formally, it will look the same, but the 3-$j$ symbols should be rewritten as Clebsch-Gordan (CG) coefficients, in bra-ket notation; this abstract notation renders them then agnostic as to dimension. In 2D, the bra-ket notated CG coefficient then becomes a Kronecker delta; in 3D, it is proportional to a 3-$j$ symbol; in 4D, it involves a Wigner 9-$j$ symbol (discussed below Eq. 17 of \cite{phil_nd}). Then the coupling coefficients may be built up as in Eq. 22 of \cite{phil_nd}; these should replace the 3-$j$ symbols in our expressions for expansions of the 3PCF, 4PCF, \textit{etc.} in the isotropic basis. 
The algorithm here could also be expanded to non-scalar fields, such as fields with spin, using the spin-weighted spherical harmonics. Since these have the same factorization properties as the harmonics used here, the extension will be straightforward.

Finally, we note that optimal sphere packing (its relevance is discussed in the next section) has also been solved in higher dimensions (for a regular packing, up to $D = 8$), so one can pack as many primaries as possible in a given volume in these extensions as well as in the 3D case.

\section{\label{sec:sphere} Using Sphere-Packing}
The optimal regular sphere packing in 3D has a volume filling factor of $f = \pi/(3\sqrt{2}) = 74.048\%$, as shown by Gauss in 1831. We can use this to compute how many non-overlapping spheres we can put in a box of volume $V = L^3$. The volume we may fill is $V_{\rm fill} = f V$, and the number of spheres is then 
\begin{align}
N_{\rm sphere} = \frac{V_{\rm fill}} {V_{\rm sphere}}
\end{align}
with 
\begin{align}
    V_{\rm sphere} = \frac{ 4\pi }{3} r_{\rm max}^3
\end{align}
and $r_{\rm max} \simeq 200$ Mpc. We then have
\begin{align}
    N_{\rm sphere} = \frac{3 f L^3}{4\pi r_{\rm max}^3};
\end{align}
for $L = 2,000$ Mpc, roughly the size of BOSS, $N_{\rm sphere} \approx 180$; for a survey of the size of DESI, roughly 30$\times$ the volume of BOSS, we have $N_{\rm sphere} \approx 5,300$. Of course mocks of any size can be used for testing a pipeline, and one just has to assess that the statistical variation of whatever quantity is of interest (\textit{e.g.} the 4PCF of the fiber-assigned field) is sufficiently small compared to some desired tolerance.

\section{\label{sec:concs} Conclusions}
We have presented an algorithm that enables realization, about a given primary galaxy, of a density field with any desired 2PCF, 3PCF, 4PCF, and even 5PCF. This should be very useful in the context of testing pipelines for upcoming NPCF analyses in large-scale structure surveys such as DESI, Euclid, Roman, Spherex, \textit{etc}. We emphasize that this algorithm produces the desired clustering \textit{about a given primary}; it can be used to produce the desired NPCFs about many such primaries, and then an algorithm to measure them, such as \textsc{encore} \cite{encore} can be run using only those primaries (the code has this functionality). While this means that the algorithm here is not truly producing a density field that has the desired NPCFs around any arbitrary point in the field, being able to produce desired NPCFs around a large number of primaries is still a significant advance useful for testing pipelines, as discussed more fully in the Introduction.

We note that the 4PCF algorithm in particular would be enabling for testing the parity-odd analysis pipeline to search for parity-violation using the galaxy 4PCF, as initially described in \cite{cahn_short} and carried out in \cite{hou_odd, phil_odd} using \cite{hou_cov} for the covariance matrix (we comment on \cite{phil_odd} in a footnote \footnote{The work of \cite{phil_odd} should unfortunately not be viewed as an independent confirmatory study, since it was done based on ideas its author took from work with our group, and the covariance matrix template was computed by a member of our group and not intended for publication. Thus this template was not fully calibrated to the final BOSS data used. This work also uses incorrect systematics weights on the galaxies. It selected a CMASS-like  sample from the BOSS combined sample (LOWZ+CMASS) by imposing a redshift cut, and thus ends up with the combined sample weights. However, those weights correct for systematics in both CMASS and LOWZ, meaning that they will impose some degree of essentially random fluctuations on the sample used (CMASS-like) from any systematic corrections required by LOWZ.}). We do note that alternative approaches exist for studying parity violation in 3D LSS, such as \cite{jeong, drew} and that since these involve correlations between non-linear functions of the density field including in some cases gradients, curls, and filtering functions, which go beyond a simple 4PCF, further work is needed to understand how they would interact with the algorithm presented here.

Finally, we note that this approach would enable producing an arbitrary CMB trispectrum; this is simply the limit of the 4PCF algorithm where $r_1 = r_2 = r_3$ identically. The approach also would work for realizing a desired set of projected statistics about a point in 2D, when binned onto annuli and on each annulus expanded in terms of Fourier modes $\exp [i m \phi]$ in the polar angle. 

\section*{Data Availability Statement}
All data used in the paper is given in the paper.

\bibliography{refs}

\end{document}